\documentclass[12pt]{article}
\usepackage{epsfig}
\def\be{\begin{equation}}
\def\ee{\end{equation}}
\def\bea{\begin{eqnarray}}
\def\eea{\end{eqnarray}}
\usepackage{graphicx}

\catcode`\@=11
\def\lsim{\mathrel{\mathpalette\@versim<}}
\def\gsim{\mathrel{\mathpalette\@versim>}}
\def\@versim#1#2{\vcenter{\offinterlineskip
\ialign{$\m@th#1\hfil##\hfil$\crcr#2\crcr\sim\crcr } }}
\catcode`\@=12

\catcode`@=12
\begin{document}
\thispagestyle{empty}
\begin{flushright}
UCRHEP-T541\\
September 2014\
\end{flushright}
\vspace{0.6in}
\begin{center}
{\LARGE \bf Anomalous Higgs Yukawa Couplings\\}
\vspace{1.2in}
{\bf Sean Fraser and Ernest Ma\\}
\vspace{0.2in}
{\sl Department of Physics and Astronomy, University of California,\\
Riverside, California 92521, USA\\}
\end{center}
\vspace{1.2in}
\begin{abstract}\
In the standard model, the Higgs boson $h$ couples to the quarks and 
charged leptons according to the well-known formula $(m_\psi/v) h \bar{\psi} 
\psi$ where $\psi =$ quark ($q$) or lepton $(l)$ and $v = 246$ GeV is its 
vacuum expectation value.  Suppose $m_\psi$ is of radiative origin instead, 
then the effective $h \bar{\psi} \psi$ Yukawa coupling will not be exactly 
$m_\psi/v$.  We show for the first time quantitatively how this may shift 
the observed branching fractions of $h \to \bar{b} b$ and 
$h \to \tau^+ \tau^-$ upward or downward.  Thus the precision measurements 
of Higgs decay to fermions at the Large Hadron Collider, due to 
resume operation in 2015, could be the key to possible new physics.  
\end{abstract}

\newpage
\baselineskip 24pt
The 2012 discovery of the 125 GeV particle~\cite{atlas12,cms12} ushered in 
a new era of particle physics.  The observed decay modes of this particle 
are consistent with it being the long sought Higgs boson $h$~\cite{e13} of 
the standard model (SM) of particle interactions.  In the near future, after 
the Large Hadron Collider (LHC) resumes operation in 2015, more 
data will allow these determinations to be greatly improved.  Of particular 
interest are the $h$ branching fractions to fermions, such as $\bar{b} b$ and 
$\tau^+ \tau^-$.  They are predicted by the SM to be proportional to 
$3m_b^2$ and $m_\tau^2$ respectively, whereas current data are not definitive 
in this regard.
In this paper, it is shown how these Yukawa couplings may be different from 
those of the SM if $m_q$ or $m_l$ is radiative in origin.  Thus the precision 
measurements of Higgs decay to fermions could be the key to possible 
new physics.

The idea that quark and lepton masses may be radiative is of course not 
new.  A review of such mechanisms already appeared 25 years 
ago~\cite{bm89}.  In the context of the SM or its left-right extension, 
early work~\cite{b88,bkm88,m89,bm90,m90,mw90,hvw90} discussed how small 
radiative masses may be generated, but did not consider how the corresponding 
Higgs Yukawa couplings are affected.  The implicit assumption is that 
the change is negligible.  In supersymmetry, in the presence of soft 
breaking in the scalar sector, there are both tree-level and one-loop 
contributions to quark masses~\cite{m89-1,mn90,bk99,chlm02}.  The resulting 
corrections to the Higgs Yukawa couplings have indeed been studied.  
Here we do something new.  We consider radiative quark or lepton mass from 
dark matter in the context of the SM with only one Higgs boson.  We show 
that the resulting Higgs Yukawa coupling may differ significantly from the 
SM prediction.

We follow the generic notion of a recent proposal which links radiative 
fermion mass with dark matter~\cite{m14,gr14}.  The first step is to forbid 
the usual Yukawa 
coupling $\bar{\psi}_L \psi_R \phi^0$ or $\bar{\psi}_L \psi_R \bar{\phi}^0$ 
by some symmetry.  The second step is to postulate new particles which 
allow this connection to be made in one loop with soft breaking of the 
assumed symmetry.  A typical realization is shown below for $m_\tau$.
\begin{figure}[htb]
\vspace*{-3cm}
\hspace*{-3cm}
\includegraphics[scale=1.0]{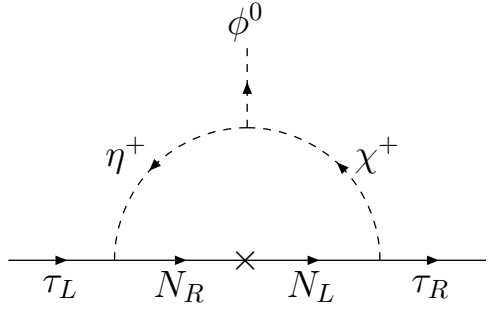}
\vspace*{-21.5cm}
\caption{One-loop generation of $m_\tau$ with $\eta^+,\chi^+$.}
\end{figure}
The new idea of this paper is the detailed analysis of the $h \bar{\psi} \psi$ 
coupling which shows for the first time that it could be significantly 
different from the SM value of $m_\psi/v$, where $v = 246$ GeV.  Note the 
important fact that this deviation comes entirely from a renormalizable 
theory.  There are no hidden assumptions and all new particles and allowed 
renormalizable interactions are considered.

In Fig.~1, $\eta^+$ is part of an electroweak doublet $(\eta^+,\eta^0)$ and 
$\chi^+$ is a singlet.    To realize this diagram in a 
simple specific model which also accommodates dark matter, consider 
the discrete symmetry $Z_2 \times Z_2$ under which $N_R, \eta, \chi$ 
are (odd, even), $N_L$ is (odd, odd), $\tau_R$ is (even, odd), and all 
other fields are (even, even).  As a result, the usual (hard) Yukawa term 
$\bar{\tau}_L \tau_R \phi^0$ is forbidden, but the (hard) 
Yukawa terms $f_\eta \bar{N}_R \tau_L \eta^+$ and $f_\chi \bar{\tau}_R N_L 
\chi^-$ are allowed, as well as  the $\mu (\eta^+ \phi^0 - \eta^0 
\phi^+) \chi^-$ trilinear interaction which mixes $\eta^\pm$ and $\chi^\pm$.
The first $Z_2$ is assumed to be exact, which accommodates dark 
matter~\cite{m06}.  The second $Z_2$ is assumed to be broken softly by 
the term $m_N \bar{N}_L N_R$.  This allows the completion of the loop. 
The resulting radiative $m_\tau$ is guaranteed to be finite and calculable 
as shown below.

The $2 \times 2$ mass-squared 
matrix spanning $(\eta^\pm,\chi^\pm)$ is given by
\begin{eqnarray}
{\cal M}^2_{\eta \chi} = \pmatrix{m^2_\eta & \mu v/\sqrt{2} \cr 
\mu v/\sqrt{2} & m^2_\chi} = \pmatrix{\cos \theta & -\sin \theta 
\cr \sin \theta & \cos \theta} \pmatrix{m_1^2 & 0 \cr 0 & m_2^2} 
\pmatrix{\cos \theta & \sin \theta \cr -\sin \theta & \cos \theta},
\end{eqnarray}
where $\langle \phi^0 \rangle = v/\sqrt{2}$ and $m^2_{\eta,\chi}$ already 
include the $\lambda_{\eta,\chi} v^2$ contributions from the quartic scalar 
terms $\lambda_\eta (\eta^\dagger \eta) (\Phi^\dagger \Phi)$ and 
$\lambda_\chi (\chi^\dagger \chi) (\Phi^\dagger \Phi)$ respectively.  Now 
\begin{equation}
\zeta_1 = \eta \cos \theta + \chi \sin \theta, ~~~ 
\zeta_2 = \chi \cos \theta - \eta \sin \theta,
\end{equation}
are the mass eigenstates, and the mixing angle $\theta$ is given by
\begin{equation}
{\mu v \over \sqrt{2}} = \sin \theta \cos \theta (m_1^2 - m_2^2).
\end{equation}
This imposes the constraint $\mu v/\sqrt{2}(m_1^2-m_2^2) < 1/2$ on 
$\mu/(m_1^2-m_2^2)$.  Together with the requirement that $m^2_{1,2} > 0$, 
this guarantees that there is no charge breaking minimum in the Higgs 
potential. The exact calculation of $m_\tau$ in terms of the exchange 
of $\zeta_{1,2}$ results in~\cite{m06,fm12,mpr13}
\begin{eqnarray}
m_\tau &=& {f_\eta f_\chi \sin \theta \cos \theta ~m_N \over 16 \pi^2} 
\left[ {m_1^2 \over m_1^2 - m_N^2} \ln {m_1^2 \over m_N^2} - 
{m_2^2 \over m_2^2 - m_N^2} \ln {m_2^2 \over m_N^2} \right] \nonumber \\ 
&=& {f_\eta f_\chi \mu  v \over 16 \sqrt{2} \pi^2 m_N} F(x_1,x_2),
\end{eqnarray}
where $x_{1,2} = m_{1,2}^2/m_N^2$ and
\begin{equation}
F(x_1,x_2) = {1 \over x_1 - x_2} \left[ {x_1 \over x_1 -1} \ln x_1 - 
{x_2 \over x_2 -1} \ln x_2 \right],~~~F(x,x) = {1 \over x-1} - 
{\ln x \over (x-1)^2}.
\end{equation}
Previous calculations usually assume that $\mu v/\sqrt{2} << m^2_{\eta,\chi}$, 
so $\eta,\chi$ are kept as mass eigenstates and $\mu v/\sqrt{2}$ as a 
coupling or mass insertion.  The presumed Higgs Yukawa coupling is 
then calculated using the same integral as that of the radiative mass, with 
their ratio unchanged from the SM prediction.  This is an unjustified 
assumption because the correct comparison is $\mu v/\sqrt{2}$ against 
$m_1^2-m_2^2$ and not $m_{1,2}^2$ as shown in Eq.~(3).

Let $\phi^0 = (v + h)/\sqrt{2}$ and consider the effective Yukawa 
coupling $h \bar{\tau} \tau$.  In the SM, it is of course equal to $m_\tau/v$, 
but here it has three contributions, one from Fig.~1 and two others 
which are new and have not been considered before.  Assuming that 
$m_h^2$ is small compared to $m_{1,2}^2$ and $m_N^2$, Fig.~1 yields 
the one-loop effective coupling $f_\tau^{(3)} h \bar{\tau} \tau$, where 
\begin{equation}
f_\tau^{(3)} = {f_\eta f_\chi \mu \over 16 \sqrt{2} \pi^2 m_N} 
[ (\cos^4 \theta + \sin^4 \theta) F(x_1,x_2) + \sin^2 \theta \cos^2 \theta 
(F(x_1,x_1) + F(x_2,x_2))].
\end{equation}
\begin{figure}[htb]
\hspace*{2cm}
\includegraphics[scale=1.5]{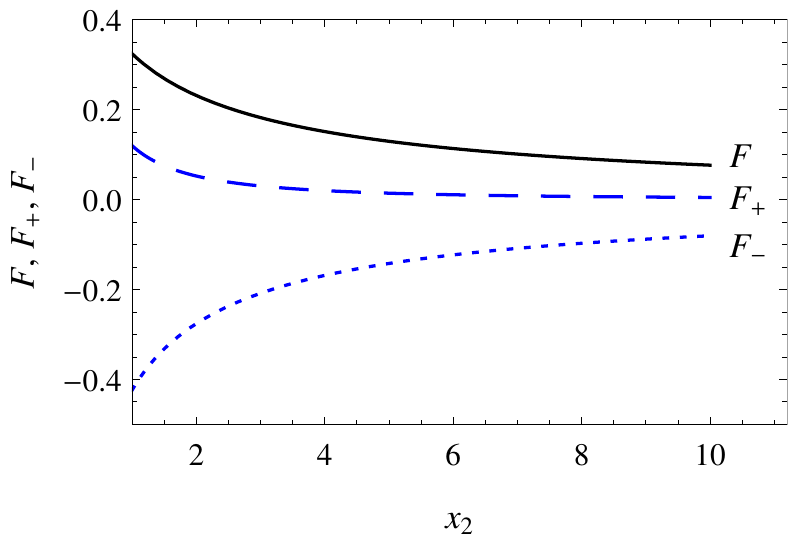}
\caption{The functions $F,F_{\pm}$ plotted against $x_2$ for $x_1=x_2+2$.}
\end{figure}
Comparing Eq.~(6) with Eq.~(4), we see that $f_\tau^{(3)} = m_\tau/v$ only 
in the limit $\sin 2 \theta \to 0$.  We see also that $F(x_1,x_1)+F(x_2,x_2)$ is 
always greater than $2F(x_1,x_2)$, so that $f_\tau^{(3)}$ is always greater 
than $m_\tau/v$.  The correction due to nonzero $m_h$ is easily computed 
in the limit $m_1=m_2=m_N$, in which case it is $m_h^2/12m_N^2$.  This 
shows that it should be generally negligible. Let
\begin{equation}
F_+(x_1,x_2) = {F(x_1,x_1) + F(x_2,x_2) \over 2 F(x_1,x_2)} - 1, ~~~ 
F_-(x_1,x_2) = {F(x_1,x_1) - F(x_2,x_2) \over 2 F(x_1,x_2)},
\end{equation}
then $F_+ \geq 0$ and if $x_1=x_2$, $F_+ = F_- = 0$.  To get an idea of 
their behavior, we take for example $x_1 = x_2 + 2$ 
and plot $F$ and $F_\pm$ as functions of $x_2$ in Fig.~2.  Note that this 
choice allows all possible $\theta$ values as long as  
$\mu v/\sqrt{2} m_N^2 = \sin 2 \theta$.

Two other one-loop contributions exist, i.e. $f_\tau^{(1,2)} h \bar{\tau} \tau$.  
They come from the quartic scalar couplings  $(\lambda_\eta/2) (v + h)^2 
\eta^+ \eta^-$ and $(\lambda_\chi/2) (v + h)^2 \chi^+ \chi^-$ respectively. 
Doing the corresponding integrals, we obtain
\begin{eqnarray} 
f^{(1)}_\tau &=& {\lambda_\eta v f_\eta f_\chi \over 
16 \pi^2 m_N} \sin \theta \cos \theta [\cos^2 \theta F(x_1,x_1)  - 
\sin^2 \theta F(x_2,x_2) - \cos 2 \theta F(x_1,x_2)], \\ 
f^{(2)}_\tau &=& {\lambda_\chi v f_\eta f_\chi \over 
16 \pi^2 m_N} \sin \theta \cos \theta [\sin^2 \theta F(x_1,x_1) 
- \cos^2 \theta F(x_2,x_2) + \cos 2 \theta  F(x_1,x_2)]. 
\end{eqnarray}
\begin{figure}[htb]
\vspace*{1cm}
\hspace*{1cm}
\includegraphics[scale=1.5]{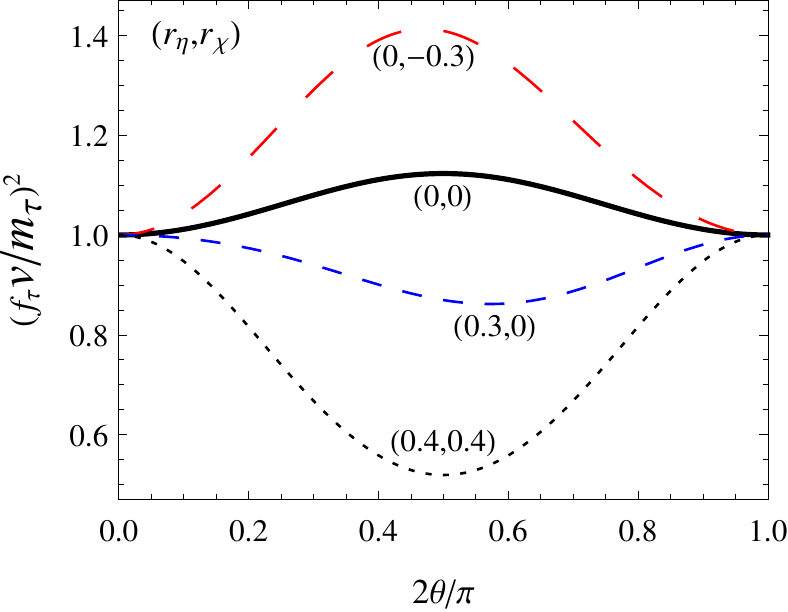}
\caption{The ratio $(f_\tau v/m_\tau)^2$ plotted against $\theta$ for 
$x_1 = 3$ and $x_2 = 1$ with various $(r_\eta,r_\chi)$.}
\end{figure}
Combining all three contributions and using Eq.~(3), we have
\begin{eqnarray}
{f_\tau v \over m_\tau} &=& {[f_\tau^{(1)} + f_\tau^{(2)} + f_\tau^{(3)}]v \over 
m_\tau} \nonumber \\ 
&=& 1 + {1 \over 2} (\sin 2 \theta)^2 \left\{ F_+ + (x_1-x_2) 
\left[ \cos 2 \theta (r_\eta - r_\chi) F_+ + 
(r_\eta + r_\chi) F_- \right] \right\},
\end{eqnarray}
where $r_{\eta,\chi} = \lambda_{\eta,\chi} (m_N/\mu)^2$.
We plot $(f_\tau v/m_\tau)^2$ or the analogous $(f_b v/m_b)^2$ 
as functions of $\theta$ in Fig.~3 for various 
$(r_{\eta},r_\chi)$ with $x_1 = 3$ and $x_2 = 1$. 
It shows that in general, it is not equal to one as the SM predicts.
The current LHC measurements of $h \to \tau^+ \tau^-$ and 
$h \to \bar{b} b$ provide the bounds 
\begin{eqnarray}
&& \left( {f_\tau v \over m_\tau} \right)^2 = 1.4 \pmatrix{+0.5 \cr -0.4}, ~~~ 
\left( {f_b v \over m_b} \right)^2 = 0.2 \pmatrix{+0.7 \cr -0.6} 
~{\rm (ATLAS)}~\cite{atlas}, \\ 
&& \left( {f_\tau v \over m_\tau} \right)^2 = 0.78 \pm 0.27, ~~~
\left( {f_b v \over m_b} \right)^2 = 1.0 \pm 0.5~~
{\rm (CMS)}~\cite{cms}.
\end{eqnarray}
To get an idea of the numbers involved, we note that $F(3,1) = 0.324$.  
Thus Eq.~(4) for $m_\tau$ yields $f_\eta f_\chi/4 \pi = 0.4 (m_N/\mu)$. 
This means that $m_N/\mu < 1$ is preferred.  On the other hand, 
Eq.~(3) yields $\sin 2 \theta = \mu v/\sqrt{2} m_N^2$.  This requires $m_N > 
174$ GeV for $m_N/\mu < 1$.  Hence a small $\theta$ requires a large 
$m_N$ as well as a large $\mu$, whereas $\theta = \pi/4$ is perfectly 
allowed for $\mu = m_N = v/\sqrt{2}$.  We note also that in models with 
two Higgs doublets and tree-level fermion couplings, radiative contributions 
to Higgs Yukawa couplings~\cite{kky14} may be significant in the case of 
$h \to \bar{b} b$ from $t$ exchange.

\begin{figure}[htb]
\vspace*{1cm}
\hspace*{1cm}
\includegraphics[scale=1.5]{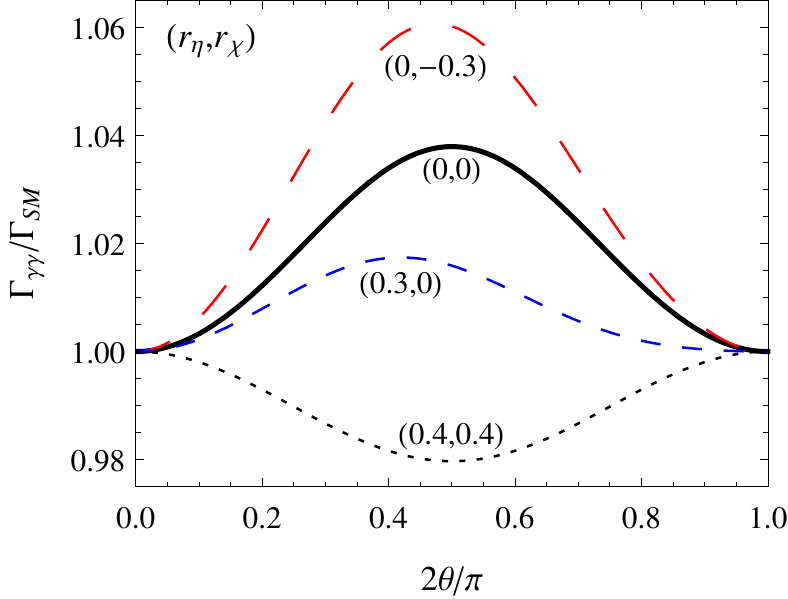}
\caption{The ratio $\Gamma_{\gamma \gamma}/\Gamma_{SM}$ plotted against 
$\theta$ for $x_1=3$ and $x_2=1$ with various $(r_\eta,r_\chi)$ and 
$\mu/m_N =1$.}
\end{figure}
The charged scalars $\zeta_{1,2}$ also contribute to 
$h \to \gamma \gamma$~\cite{abg12,sk13}.  Its decay rate is given by 
\begin{equation}
\Gamma_{\gamma \gamma} = {G_F \alpha^2 m_h^3 \over 128 \sqrt{2} \pi^3} 
\left| {4 \over 3} 
A_{1/2} \left( {4 m_t^2 \over m_h^2} \right) + 
A_{1} \left( {4 M_W^2 \over m_h^2} \right) + 
f_1 A_{0} \left( {4 m_1^2 \over m_h^2} \right) + 
f_2 A_{0} \left( {4 m_2^2 \over m_h^2} \right) \right|^2,
\end{equation}
where
\begin{eqnarray}
f_1 &=& {1 \over 4 x_1} (\sin 2 \theta)^2 (x_1-x_2) \left\{ 1 + 
{1 \over 2} (x_1-x_2) [ (r_\eta + r_\chi) + \cos 2 \theta (r_\eta - r_\chi)] 
\right\}, \\  
f_2 &=& {1 \over 4 x_2} (\sin 2 \theta)^2 (x_1-x_2) \left\{ -1 + 
{1 \over 2} (x_1-x_2) [ (r_\eta + r_\chi) - \cos 2 \theta (r_\eta - r_\chi)] 
\right\}.
\end{eqnarray}
The $A$ functions are well known, i.e.
\begin{eqnarray}
A_0 (y) &=& - y [1-y f(y)], \\ 
A_{1/2} (y) &=& 2 y [1 + (1-y) f(y)], \\ 
A_1 (y) &=& - [2 + 3y + 3y(2-y) f(y)],
\end{eqnarray}
where $f(y) = \arcsin^2 (y^{-1/2})$ for $y \geq 1$.  We plot in Fig.~4 the 
ratio $\Gamma_{\gamma \gamma}/\Gamma_{SM}$ as a function of $\theta$ for 
various values of $(r_\eta,r_\chi)$ with $x_1 = 3$ and $x_2 =1$, assuming also 
$\mu/m_N = 1$, i.e. $m_N \sin 2 \theta = v/\sqrt{2}$.  
The current LHC measurements of 
$h \to \gamma \gamma$ provide the bounds 
\begin{eqnarray}
&& {\Gamma_{\gamma \gamma} \over \Gamma_{SM}} = 1.57 
\pmatrix{+0.33 \cr -0.28}~{\rm (ATLAS)}, \\
&& {\Gamma_{\gamma \gamma} \over \Gamma_{SM}} = 0.78 \pm 0.27~{\rm (CMS)}.
\end{eqnarray}

\begin{figure}[htb]
\vspace*{1cm}
\hspace*{1cm}
\includegraphics[scale=1.5]{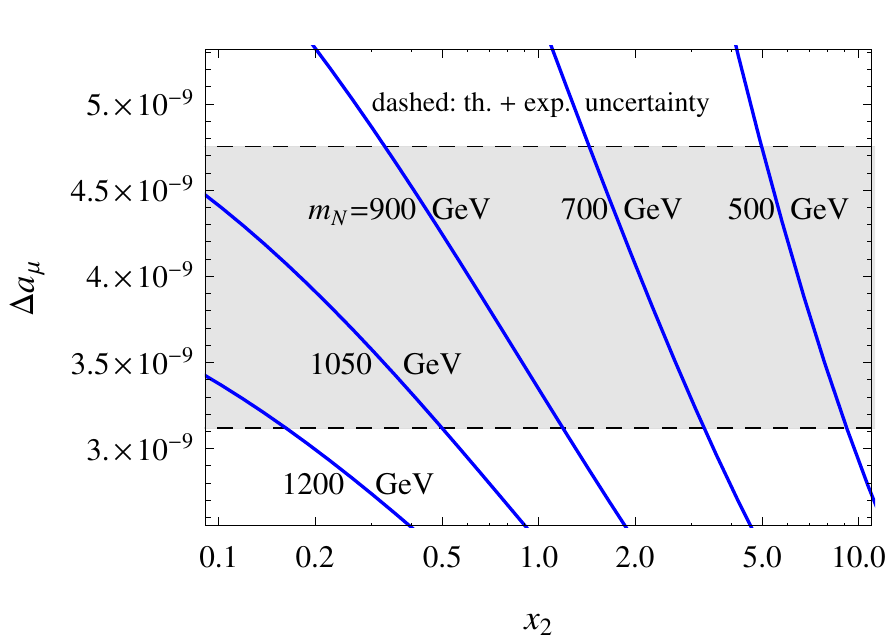}
\caption{$\Delta a_\mu$ plotted against $x_2$ with $x_1 = x_2 +2$ 
for various $m_N$.}
\end{figure}
Another important consequence of the radiative generation of fermion masses 
is the induced electromagnetic interaction which is now related to the 
fermion mass itself.  If we apply the above procedure to the muon (using a 
different $N$ and thus also different $f_{\eta,\chi}$), we find 
its anomalous magnetic moment to be given by
\begin{equation}
\Delta a_\mu = {(g-2)_\mu \over 2} = {m_\mu^2 \over m_N^2} \left[ 
{G(x_1) - G(x_2) \over H(x_1) - H(x_2)} \right],
\end{equation}
where
\begin{equation}
G(x) = {2 x \ln x \over (x-1)^3} - {x + 1 \over (x-1)^2}, ~~~ 
H(x) = {x \ln x \over x-1}.
\end{equation}
The current discrepancy of the experimental measurement~\cite{g-2} versus 
the most recently updated theoretical calculation~\cite{bddj12} is
\begin{equation}
\Delta a_\mu = a_\mu^{\rm exp} - a_\mu^{\rm SM} = 39.35 \pm 5.21_{\rm th} 
\pm 6.3_{\rm exp} \times 10^{-10}.
\end{equation}
We plot in Fig.~5 this theoretical prediction for various $m_N$ as a 
function of $x_2$ with $x_1 = x_2 + 2$, i.e. $\sin 2 \theta = 
\mu v/\sqrt{2} m_N^2$.  Also shown is 
Eq.~(23) with the experimental and theoretical uncertainties combined 
in quadrature.  We note that the maximum value of $\Delta a_\mu$ 
is obtained in the limit of $x_1 = x_2 = 1$ where 
$\Delta a_\mu = m_\mu^2/3m_N^2$. 
We note also that the subdominant contributions to $\Delta a_\mu$ from 
$f_\eta^2$ and $f_\chi^2$ are negative as expected~\cite{kt12}, i.e.
\begin{eqnarray}
(\Delta a_\mu)' = {-m_\mu^2 \over 16 \pi^2 m_N^2} \left\{ f_\eta^2  
 \left[ \cos^2 \theta J(x_1) + \sin^2 \theta J(x_2) \right] 
+  f_\chi^2 
 \left[ \sin^2 \theta J(x_1) + \cos^2 \theta J(x_2) \right] \right\},
\end{eqnarray}
where
\begin{equation}
J(x) = { x \ln x \over (x-1)^4} + {x^2 - 5x - 2 \over 6(x-1)^3}.
\end{equation}
Using Eq.~(4) for $m_\mu$, we have checked that the value of 
$(f_\eta f_\chi/4 \pi) (\mu/m_N)$ varies between 0.01 and 0.1 in this 
range.  If $m_\mu$ and $m_e$ are both radiative, then $\mu \to e \gamma$ 
would be severely constrained~\cite{mnw90}.  However, a flavor symmetry 
such as $Z_3$ may exist~\cite{m10} to forbid it.

\begin{figure}[htb]
\vspace*{-3cm}
\hspace*{-3cm}
\includegraphics[scale=1.0]{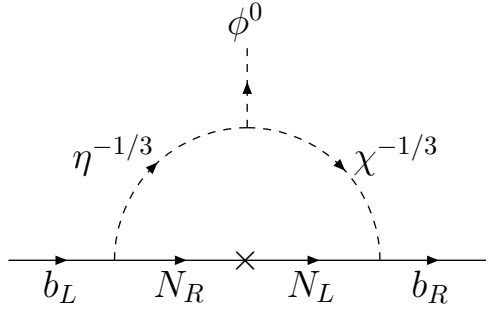}
\vspace*{-21.5cm}
\caption{One-loop generation of $m_b$ with $\eta^{-1/3},\chi^{-1/3}$.}
\end{figure}
As for quarks, radiative $m_t$ would require very large corresponding 
Yukawa couplings in the loop.  This is perhaps unrealistic, but such 
is not an issue with the other quarks.   If the $b$ quark mass is radiative 
as proposed in Ref.~\cite{m14}, the one-loop diagram is given by Fig.~6.
Here there are colored scalar triplets: $(\eta^{2/3},\eta^{-1/3})$, $\chi^{-1/3}$, 
which are respectively doublet and singlet under $SU(2)$.  The decay rate 
of $h \to g g$ is then modified.
\begin{equation}
\Gamma_{g g} = {G_F \alpha_S^2 m_h^3 \over 64 \sqrt{2} \pi^3} 
\left| A_{1/2} \left( {4 m_t^2 \over m_h^2} \right) +  
\left[ f'_1 A_{0} \left( {4 {m'_1}^2 \over m_h^2} \right) + 
f'_2 A_{0} \left( {4 {m'_2}^2 \over m_h^2} \right) + 
f'_\eta A_{0} \left( {4 {m'_\eta}^2 \over m_h^2} \right) \right] \right|^2.
\end{equation}
Here $m'_{1,2}$ refer to the mass eigenvalues of the mixed 
$(\eta^{-1/3}, \chi^{-1/3})$ system and $m'_\eta$ is the mass of $\eta^{2/3}$ 
with ${m'_\eta}^2 = {m'_1}^2 \cos^2 \theta' + {m'_2}^2 \sin^2 \theta'$. 
Using $r'_{\eta,\chi} = \lambda'_{\eta,\chi} (m_N/\mu')^2$ and $x'_{1,2,\eta} = 
{m'_{1,2,\eta}}^2/m_N^2$, we find
\begin{figure}[ht]
\vspace*{1cm}
\hspace*{1cm}
\includegraphics[scale=1.5]{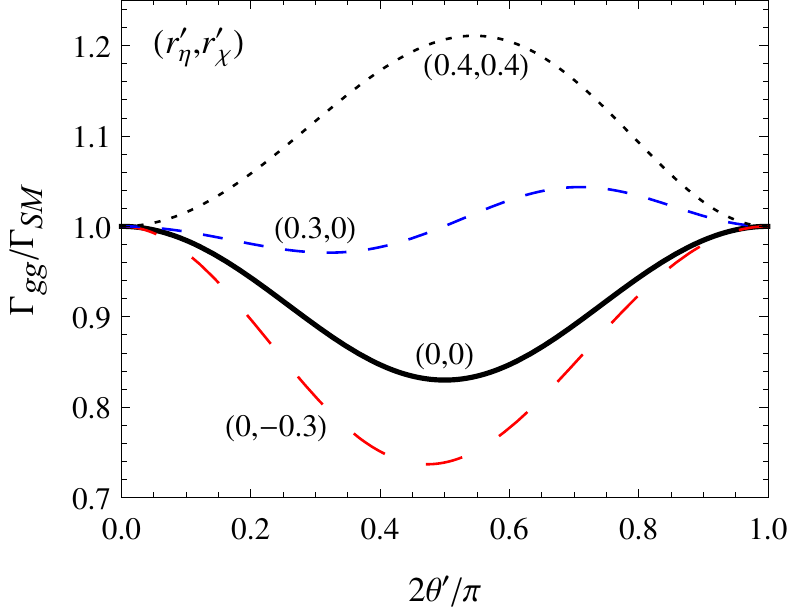}
\caption{The ratio $\Gamma_{g g}/\Gamma_{SM}$ plotted against $\theta'$ for 
$x'_1=3$ and $x'_2=1$ with various $(r'_\eta,r'_\chi)$ and $\mu'/m_N = 1$.}
\end{figure}
\begin{eqnarray}
f'_1 &=& {1 \over 4 x'_1} (\sin 2 \theta')^2 (x'_1-x'_2) \left\{ 1 + 
{1 \over 2} (x'_1-x'_2) [ (r'_\eta + r'_\chi) + \cos 2 \theta' (r'_\eta - r'_\chi)] 
\right\}, \\  
f'_2 &=& {1 \over 4 x'_2} (\sin 2 \theta')^2 (x'_1-x'_2) \left\{ -1 + 
{1 \over 2} (x'_1-x'_2) [ (r'_\eta + r'_\chi) - \cos 2 \theta' (r'_\eta - r'_\chi)] 
\right\}, \\ 
f'_\eta &=& {r'_\eta \over 4 x'_\eta} (\sin 2 \theta')^2 (x'_1 - x'_2)^2.
\end{eqnarray}
We plot in Fig.~7 the ratio $\Gamma_{g g}/\Gamma_{SM}$ as a function of 
$\theta'$ for $x'_1 = 3$ and $x'_2=1$ with $\mu'/m_N = 1$ for various 
$(r'_\eta,r'_\chi)$. 
This shows that the production of $h$ via gluon fusion may be significantly 
affected.  Currently the sample of $h \to \bar{b} b$ decays at the LHC 
comes only from vector boson associated production, which is unchanged 
in this case.

In conclusion, we have shown in this paper in detail how the 
fermionic decay of the 125 GeV particle $h$ discovered at the LHC in 2012 
may differ from the expectations of the standard model if a quark or lepton  
$\psi$ acquires its mass radiatively.  The Yukawa coupling of $h$ to 
$\bar{\psi} \psi$ is then predicted to differ in general from the 
standard-model prediction of $m_\psi/v$ where $v=246$ GeV.  A large effect 
is possible with a modest mixing angle defined in Eq.~(3).  This deviation 
is possibly observable in $h \to \tau^+ \tau^-$ and $h \to \bar{b} b$ 
with more data at the LHC, due to resume operation in 2015.  The new 
particles responsible for this deviation are different from other 
possible explanations of this effect, and if observed, could help 
to distinguish our proposal from others.

\noindent \underline{Acknowledgment}~:~This work is supported in part 
by the U.~S.~Department of Energy under Grant No.~DE-SC0008541.

\bibliographystyle{unsrt}

\end{document}